\documentclass[11pt,letterpaper]{article}
\pdfoutput=1
\usepackage{jheppub}
\usepackage{color,float,graphicx,hyperref}
\usepackage{amsmath,amssymb,subfig}
\usepackage{soul}

\newcommand{\cO}{\mathcal{O}}
\newcommand{\qqvh}{q\bar{q}\kern0.1em V\kern-0.1em H}

\begin{document}

\title{Higgs to $b\bar{b}$ from Vector Boson Fusion for High-Scale Physics} 
\author{Tao Han,}
\author{Sze Ching Iris Leung,}
\author{Matthew Low}

\affiliation{PITT PACC, Department of Physics and Astronomy, University of Pittsburgh, 3941 O’Hara St., Pittsburgh, PA 15260, USA}

\emailAdd{than@pitt.edu}
\emailAdd{szl13@pitt.edu}
\emailAdd{mal431@pitt.edu}
\date{\today}

\abstract{Vector boson fusion is arguably the most direct collider probe of electroweak symmetry breaking.  Typically, the signature includes two forward/backward jets with low transverse momenta with a  scale that is set by the mass of the vector boson.  For this reason, an upper cut is used when searching for vector boson fusion processes in the Standard Model.  Alternatively, the upper cut on the forward jets can be removed and the high-momentum exchange region of vector boson fusion can be studied.  This phase space region has sensitivity to new physics via higher dimensional operators and form factors.  In this work, we study the high-momentum region of the vector boson fusion channel where the Higgs decays to $b\bar{b}$.  We show that, depending on the form of new physics, the limits on the new physics scale range from $0.5$ TeV to $1.8$ TeV.}

\maketitle

\section{Introduction}
\label{sec:introduction}

The milestone discovery of the Higgs boson $(h)$ by the ATLAS~\cite{ATLAS:2012yve} and CMS~\cite{CMS:2012qbp} collaborations at the CERN Large Hadron Collider (LHC) is of great theoretical and experimental significance in understanding our microscopic world. The Standard Model (SM) of the strong and electroweak interactions, along with the Higgs mechanism, provides us with a consistent theoretical framework valid up to high scales.  On the other hand, we are still lacking an understanding of if and how the Higgs mass, the only dimensionful parameter in the theory, is stabilized against quantum corrections from a higher new physics scale.  This is the electroweak hierarchy problem~\cite{Weinberg:1975gm,Gildener:1976ai,tHooft:1980xss}.  It is thus of fundamental importance to experimentally probe the interactions of the Higgs at higher energy scales, which should hold the key to understanding the hierarchy problem. 

Some efforts have been made along these lines to probe the largest SM Higgs coupling $ht\bar{t}$ at the LHC in high scales via the leading Higgs production channel of gluon fusion $gg\to h\to ZZ$~\cite{Goncalves:2020vyn} and associated production $pp\to t\bar{t} h$~\cite{MammenAbraham:2021ssc}.  The next leading production mechanism of the Higgs boson is the process by which two vector bosons collide to produce a Higgs boson, known as vector boson fusion (VBF).  With the Higgs coupling directly proportional to the gauge boson mass,  this process is one of the most direct probes of electroweak symmetry breaking that we have.  It is well-known that without a Higgs boson, the amplitude for longitudinally-polarized gauge boson scattering exhibits quadratic growth with energy, violating perturbative unitarity at a scale of $\Lambda \approx 2.2~{\rm TeV}$~\cite{Lee:1977eg,Lee:1977yc}.  This is cornerstone of the no-lose theorem of the LHC uncovering the first step of electroweak symmetry breaking~\cite{Glashow:1961tr,Salam:1964ry,Weinberg:1967tq}.

The traditional VBF signal leverages the fact that the two outgoing quarks exchange a relatively low amount of momentum with the vector bosons.  These quarks, often called tagging jets \cite{Kleiss:1987cj,Barger:1988mr} tend to have high pseudorapidity $(\eta)$ and low transverse momentum $(p_T)$, dictated by the (collinear) $W$ propagator (see Refs.~\cite{Rauch:2016pai,Green:2016trm,Anders:2018oin,Bellan:2019xpr,Baglio:2020bnc,Covarelli:2021gyz,Chen:2016wkt} for reviews).   Signal regions are then defined by a lower cut (to pass trigger thresholds) and an upper cut (to maximize the VBF sensitivity) on the tagging jet $p_T$.  As expected, VBF processes have contributed to the Higgs boson discovery~\cite{ATLAS:2012qaq,CMS:2012zhx}.  In the hope to probe Higgs physics at a high scale, an alternative approach is to examine the Higgs boson with a higher transverse  momentum. This signal region, by contrast, would look to isolate events with a large momentum exchange between the vector bosons and the quarks.  At a cost of lower signal event rate, we may have increased sensitivity to physics beyond the Standard Model (BSM).

The idea of probing high energy tails of distributions at a hadron collider, in the search for higher dimensional operators, was elucidated in Drell-Yan~\cite{Farina:2016rws} and subsequently applied to diboson production~\cite{Franceschini:2017xkh}.  Searches for higher dimensional operators in VBF have been initiated in a few studies~\cite{Banerjee:2013apa,Gomez-Ambrosio:2018pnl,Biekotter:2020flu,Araz:2020zyh,Ethier:2021ydt,Biswas:2021qaf,Hwang:2023wad}.  Ref~\cite{Araz:2020zyh}, in particular, applied the techniques from~\cite{Franceschini:2017xkh} to VBF production for the $h\to\gamma\gamma$ and $h\to\tau^+\tau^-$ decay channels.  They found with $3~{\rm ab}^{-1}$ a limit of $\Lambda > 3.9~{\rm TeV}$ (at 68\% C.L.) could be set on the scale of new physics.

\begin{figure}  [H]
  \begin{center}
  \includegraphics[width=1.0\textwidth]{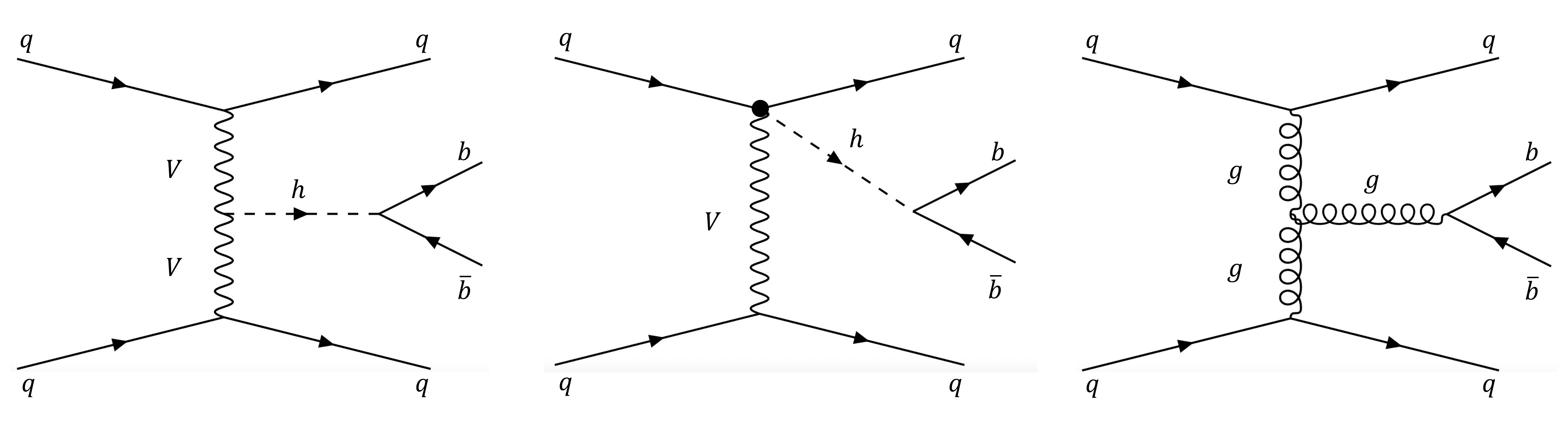}
  \caption{Representative Feynman diagrams of vector boson fusion in the Standard Model (left), via a contact interaction $\qqvh$ from dimension-6 operators (center), and the QCD background (right).}
  \label{fig:feynmanDiagram}
  \end{center}
\end{figure}

In this work, we first compute the sensitivity to new physics, parametrized by the leading dimension-6 operators and a non-local momentum-dependent form factor, using the $b\bar{b}$ decay of the Higgs, including detector simulation and backgrounds.  A sample Feynman diagram for this process is shown in Fig.~\ref{fig:feynmanDiagram}.  We find that the sensitivity is slightly worse than in the $h\to\gamma\gamma$ channel, primarily due to the fact that the Higgs mass resolution is reduced and the backgrounds are larger. Combining multiple decay channels of the Higgs would further improve the overall sensitivity. 

Secondly, we point out that the high momentum Higgs region can be effectively isolated by cutting on the tagging jets.  In channels where the Higgs four-momentum can be fully reconstructed, like in $h\to\gamma\gamma$, the $p_T$ of the Higgs is the best variable.  However, in channels where that is not possible, like in $h\to W(\ell\nu)W(\ell\nu)$, cutting on the tagging jets only loses a little sensitivity.

The outline of the paper is as follows.  In Sec.~\ref{sec:newphysics} we describe the parametrizations we use for new physics.  Section~\ref{sec:kinematics} studies the kinematics of this topology and compares a few approximations.  Next, we discuss the details of our simulation in Sec.~\ref{sec:simulation}.  The results are shown in Sec.~\ref{sec:results} and we present our conclusions in Sec.~\ref{sec:conclusions}.

\section{New Physics}
\label{sec:newphysics}

In this section, we describe the parametrizations we use for exploring new physics associated with Higgs production at high scales.

\subsection{Higher-Dimensional Operators}

A consistent and convenient scheme to parametrize new physics effects from a higher scale is the effective field theory (EFT) approach with higher dimensional operators.  The leading operators are at dimension-6.  We follow the usual parametrization of
\begin{equation}
\Delta \mathcal{L}_{d=6} = \sum_i \frac{c_i}{\Lambda^2} \cO_i .
\end{equation}
The coefficients $c_i$ are dimensionless effective couplings and $\Lambda$ is the scale of new physics. From an EFT perspective the coefficients $c_i$ could naturally be of the size $g^2$ or $\cO(1)$ (where $g$ is some coupling), in weakly-coupled BSM theories~\cite{Giudice:2007fh}, or any values $\lesssim (4\pi)^2$~\cite{Manohar:1983md,Georgi:1984zwz}  for a strongly-coupled theory. They could 
have either sign. 
Although coefficients as large as $c_i \sim (4\pi)^2$ are technically possible, it requires the unlikely scenario of strong coupling between new physics and all particles of the $\qqvh$ interaction.  Even a coupling of $c_i \sim 4\pi$ is not likely~\cite{Liu:2016idz,Franceschini:2017xkh}.

In VBF, our interaction of interest is $\qqvh$, for which there are four leading operators~\cite{Franceschini:2017xkh}. These are operators leading in the energy growth of the SM-BSM interference term. In the Warsaw basis~\cite{Grzadkowski:2010es}, they are written as $\cO_Q^{(3)}$, $\cO_Q$, $\cO^u_R$, and $\cO^d_R$
\begin{subequations} \label{eq:ops}
\begin{align}
\cO_Q^{(3)} &= (\bar{Q} \sigma^a \gamma^\mu Q) (iH^\dagger \sigma^a \overleftrightarrow{D_\mu} H), \\
\cO_Q &= (\bar{Q} \gamma^\mu Q) (iH^\dagger \overleftrightarrow{D_\mu} H), \\
\cO^u_R &= (\bar{u_R} \gamma^\mu u_R) (iH^\dagger \overleftrightarrow{D_\mu} H), \\
\cO^d_R &= (\bar{d_R} \gamma^\mu d_R) (iH^\dagger \overleftrightarrow{D_\mu} H),
\end{align}
\end{subequations}
and have been labelled as the high energy primaries~\cite{Franceschini:2017xkh}.  The explicit contribution is depicted by the diagram in Fig.~\ref{fig:feynmanDiagram} (center).

In searches for BSM physics, the largest contribution usually comes from the interference between the SM amplitude and the BSM amplitude, as the leading power correction. The ideal situation for maximal sensitivity is for the BSM amplitude to grow with energy while the SM amplitude remains constant.  For $\qqvh$ this is the case for the operators in Eq.~\eqref{eq:ops}~\cite{Franceschini:2017xkh,Araz:2020zyh} 
because they are sensitive to the longitudinal polarization of the vector boson~\cite{Azatov:2019xxn}.

One subleading operator that can be considered is
\begin{equation} \label{eq:opw}
\cO_{\phi W} = H^\dagger H W^a_{\mu\nu} W^{a,\mu\nu}.
\end{equation}
This operator is sensitive to the transverse polarizations of the vector boson and has a smaller interference term with the SM.  In the case of diboson, the sensitivity to $c_{\phi W}$ is about an order of magnitude worse than to $c^{(3)}_Q$~\cite{Bishara:2020vix}.  The coefficient $c_{\phi W}$ can technically saturate perturbativity despite the fact that this does not occur in common BSM theories.  Its expected size is actually loop suppressed and is $c_{\phi W} \sim g^4 / (16\pi^2)^2$ when the new physics in the loop is weakly-coupled or $c_{\phi W} \sim g^2 / (16\pi^2)$ when the new physics in the loop is strongly-coupled~\cite{Giudice:2007fh,Bishara:2020vix}.

For VBF production, Ref.~\cite{Araz:2020zyh} found that combining the $h\to\gamma\gamma$ and $h\to\tau^+\tau^-$ decay channels resulted in a sensitivity to $\Lambda \approx 3.9$ TeV for $3~{\rm ab}^{-1}$ at 68\% C.L.  The same operators have been constrained in diboson searches in the channels: $Wh$, $Zh$, $WZ$, and $WW$~\cite{Ellis:2014dva,Franceschini:2017xkh,Liu:2018pkg,Banerjee:2018bio,Grojean:2018dqj,Azatov:2019xxn,Freitas:2019hbk,Banerjee:2019pks,Brehmer:2019gmn,Chiu:2019ksm,Banerjee:2019twi,Baglio:2020oqu,Bishara:2020vix,Bishara:2020pfx,Huang:2020iya,Bishara:2022vsc}.  For the same amount of data, but at 95\% C.L., the limits extend to $\Lambda \approx 10~{\rm TeV}$ in the diboson channel~\cite{Franceschini:2017xkh}.  VBF and diboson production probe a different linear combination of the operators in Eq.~\eqref{eq:ops} so measuring both channels is essential to measure the full operator dependence. 

\subsection{Form Factors}

If the underlying mechanism for electroweak symmetry breaking is due to some new strong dynamics, it is conceivable to consider that the Higgs boson may not be fundamental, but a composite particle arising from strongly interacting new dynamics at a scale $\Lambda$~\cite{Panico:2015jxa}.  In such scenarios, the Higgs interaction may exhibit a momentum-dependent form-factor near or above the new physics scale $\Lambda$, rather than a point-like interaction. 

It is challenging to write a form-factor, in a general form, without prior knowledge of the strong dynamics of the specific composite scenario. Inspired by the nucleon form-factor~\cite{Punjabi:2015bba}, we adopt the following phenomenological ansatz
by adding a form factor to the $hW^+W^-$ and $hZZ$ vertices 
\begin{subequations}
    \begin{align}
(g m_W) h W^+ W^- & \to \Gamma(Q^2; n, \Lambda^2) \; (g m_W) h W^+ W^- , \\ 
\left(\frac{g m_Z}{2 c_W}\right) h Z Z & \to \Gamma(Q^2; n, \Lambda^2) \; \left(\frac{g m_Z}{2 c_W}\right) h Z Z , 
\end{align}\end{subequations}
where $c_W$ is the cosine of the weak mixing angle and the form factor is of the form~\cite{Goncalves:2020vyn,MammenAbraham:2021ssc} 
\begin{equation} \label{eq:formFactor}
\Gamma(Q^2 ; n, \Lambda^2) = \frac{1}{(1 + Q^2 / \Lambda^2)^n}.
\end{equation}
This form factor approaches 1 at low energies as $Q^2/\Lambda^2 \to 0$, and suppresses this interaction as $Q^2/\Lambda^2$ grows at high energies.  The suppression depends on the value of $n$ chosen; the larger $n$ is, the faster the suppression. The case of $n=2$ corresponds to a dipole form factor, as often adopted for the nucleon form factor.  Limits from a recent ATLAS search can be applied to this form factor for the $ht\bar{t}$ vertex~\cite{ATLAS:2023dnm}.

In our study we use $Q=p_T^h$ since the transverse momentum of the Higgs characterizes the momentum exchange.  This also ensures that Eq.~\eqref{eq:formFactor} always results in a suppression, which is the expected behavior given considerations to the hierarchy problem.

Another option, which we do not pursue in this study, is to use form factors computed from integrating out heavy resonances~\cite{Bittar:2022wgb}.  This would be useful for a future study since these form factors could be matched directly to known BSM theories.

\section{Kinematics}
\label{sec:kinematics}

It is important to explore the kinematics of the Higgs boson produced in vector boson fusion in the hope to optimize the dynamical effects. The main purpose is to identify to what extent the phase space with a boosted Higgs can be isolated purely from the kinematics of the two tagging jets.\footnote{In the $b\bar{b}$ channel that we study here, as well as in the $\gamma\gamma$ and hadronic $\tau^+\tau^-$ channels~\cite{Araz:2020zyh} the Higgs is fully reconstructable so one can always directly use $p_T^h$.  Using the tagging jets would be useful in other channels like $W(\ell\nu)W(\ell\nu)$.}

At leading order there are three objects in the final state: the Higgs $h$ and two tagging jets $j_1$ and $j_2$ (where $p_T^{j_1} \geq p_T^{j_2}$ by convention).  Balancing transverse momentum leads to the relationship
\begin{equation} \label{eq:kinematics}
(\tilde{p}_T^h)^2 = (p_T^{j_1})^2 + (p_T^{j_2})^2 + 2 p_T^{j_1} p_T^{j_2} \cos(\phi^{j_1} - \phi^{j_2}),
 \end{equation}
 where $\phi^x$ is the azimuthal angle of object $x$.  The notation $\tilde{p}_T^h$ is to indicate that this would be the transverse momentum of the Higgs in the three-body limit.

We are interested in the kinematical region with a high $p_T$ Higgs. 
There are two notable limits of Eq.~\eqref{eq:kinematics}.  The first is when $p_T^{j_2} \ll p_T^{j_1}, p_T^{h}$.  In this case the transverse momentum of the Higgs is primarily balanced by the hardest jet and the three-body space approximately collapses to a two-body space.  Even at the high luminosity LHC, this kinematic configuration is rather rare and out of reach.  It could be relevant at a future collider.

\begin{figure} [H]
  \begin{center}
  \includegraphics[width=0.45\textwidth]{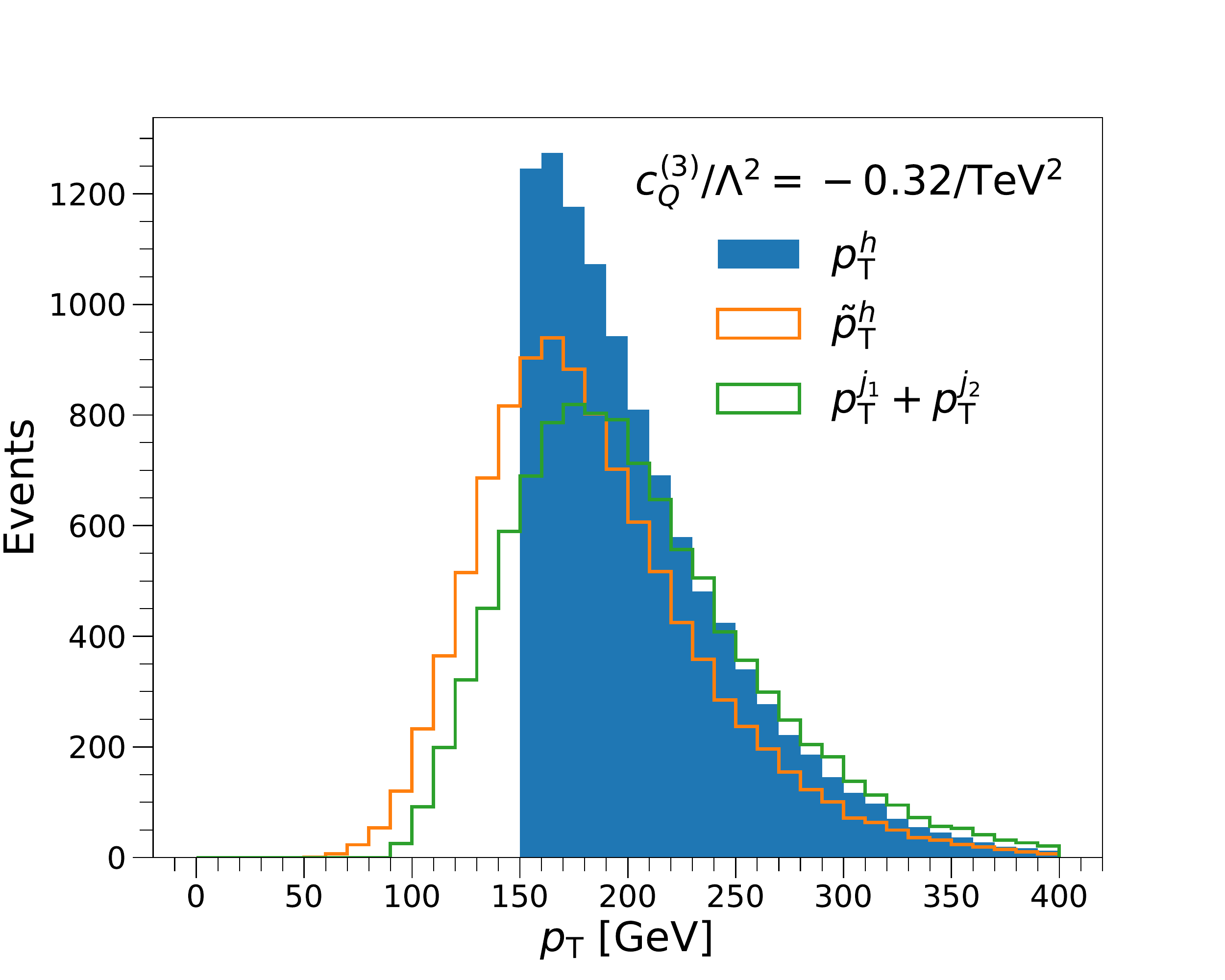} \quad\quad
  \includegraphics[width=0.45\textwidth]{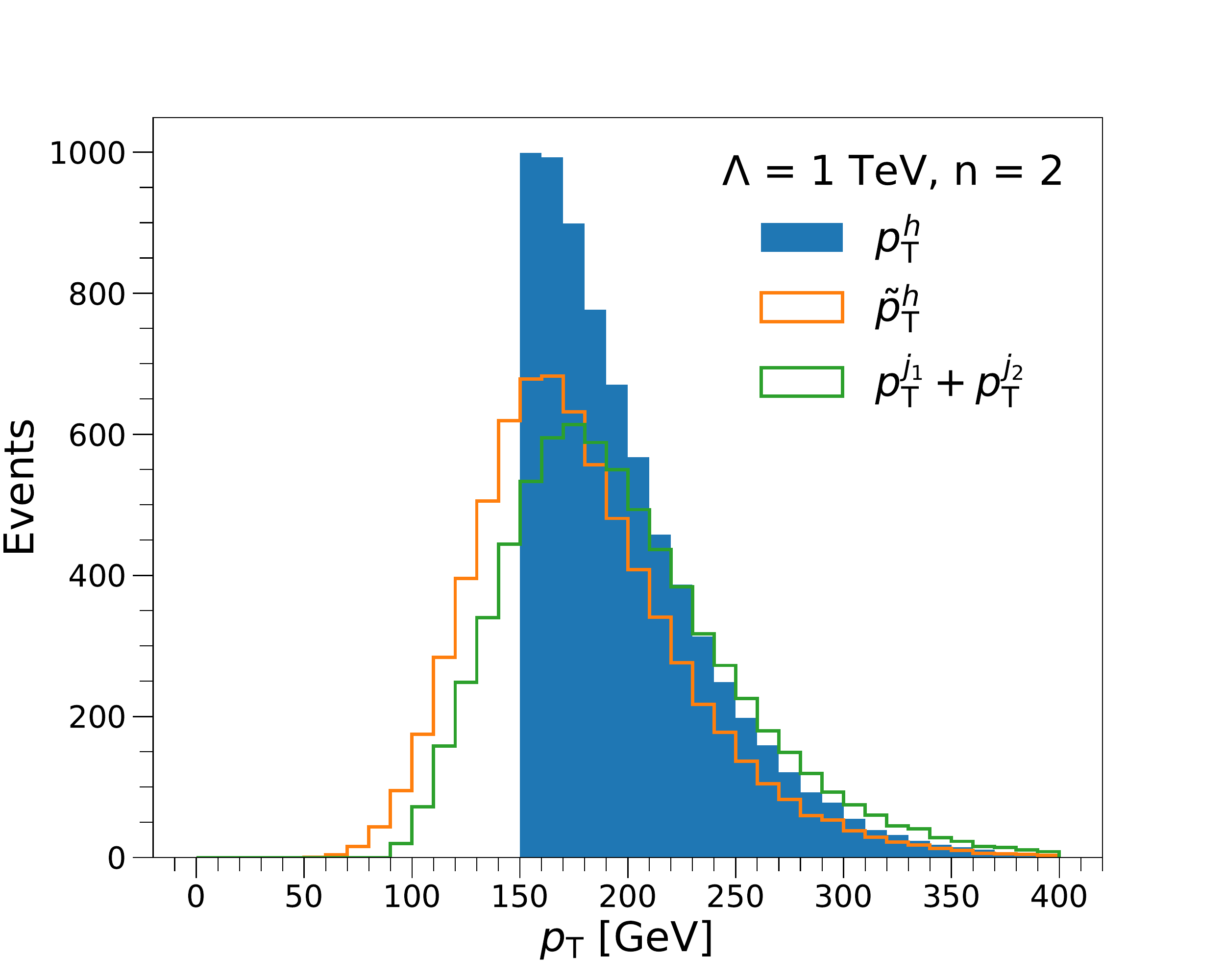}
  \caption{Distribution of Higgs $p_T$-like variables for operator $\cO_Q^{(3)}$ (left) and for the $n=2$ form factor (right) with $\mathcal{L} = 3~{\rm ab}^{-1}$. }
  \label{fig:kinematics}
  \end{center}
\end{figure}

The second is when $\cos(\phi^{j_1} - \phi^{j_2}) = \cO(1)$.  Then $p_T^h \approx p_T^{j_1} + p_T^{j_2}$.  In Fig.~\ref{fig:kinematics} we compare this quantity with $p_T^{h}$ and with $\tilde{p}_T^h$ from Eq.~\eqref{eq:kinematics}. The events are generated using {\tt Pythia 8.302}~\cite{Sjostrand:2014zea} with $p_T^h \geq 150~{\rm GeV}$.  The quantity $\tilde{p}_T^h$ is generally an underestimate due to additional radiation that is ignored in the three-body limit.  Empirically, $p_T^{j_1} + p_T^{j_2}$ is a reasonable proxy for $p_T^h$ at high values of $p_T^h$.

A number of other Higgs $p_T$-like variables can be studied.  These include $p_T^{j_1}$, $p_T^{j_2}$, and the geometric mean $\sqrt{p_T^{j_1} \; p_T^{j_2}}$ which was suggested in Refs.~\cite{Banerjee:2013apa,Biswas:2021qaf}.  We find that $p_T^h$ gives the best results, followed closely by the sum $p_T^{j_1} + p_T^{j_2}$ and the quantity $\tilde{p}_T^h$.  The other Higgs $p_T$-like variables are less correlated with $p_T^h$.

\section{Simulation}
\label{sec:simulation}

We generate events using {\tt MadGraph 5} v2.7.3~\cite{Alwall:2011uj}.  Both signal and background are generated at leading order at a center-of-mass energy $\sqrt{s} = 13$ TeV with the parton distribution function NNPDF23\textunderscore lo\textunderscore as\textunderscore0130\textunderscore qed~\cite{Ball:2013hta} through LHAPDF~\cite{Buckley:2014ana}.  Events are showered and hadronized with {\tt Pythia 8.302}, and detector simulation is performed by {\tt Delphes 3.4.1}~\cite{deFavereau:2013fsa}. Jets are reconstructed with the anti-$k_T$ algorithm with radius $R = 0.5$~\cite{Cacciari:2008gp}.  For $b$-tagging we use the truth flavor from {\tt Delphes} and apply a flat $b$-tagging efficiency of $77\%$ for the leading $b$-jet and $85\%$ for the subleading $b$-jet.\footnote{This means for events with the leading truth $b$-jet with $p_T>85~{\rm GeV}$ and the subleading truth $b$-jet with $p_T>65~{\rm GeV}$, 65\% of them pass the $b$-jet cuts ($0.77 \times 0.85 = 0.65$).}

Our signal is the process $pp \to h(b\bar{b})jj$ with either the dimension-6 operators or the form factor turned on. Our calculation for this process includes the contributions from the Standard Model, new physics, and the interference between these two.  We label the signal as {\tt BSM}.  We use dimension-6 operators from the {\tt SMEFTatNLO} model~\cite{Degrande:2020evl} via FeynRules~\cite{Alloul:2013bka}.

There are two dominant backgrounds: the Standard Model VBF contribution and the Standard Model QCD contribution.   Standard Model VBF includes a resonant Higgs and is generated from $pp \to h(b\bar{b})jj$ with all new physics contributions set to zero.  
We label this background as {\tt VBF}.  The QCD background is generated by $pp \to b\bar{b}jj$ (excluding on-shell Higgs contributions).  We label this background as {\tt QCD}.

\subsection*{Pre-Selection and Validation}

We apply cuts to the detector-level events in two stages.  The first is the pre-selection which we use for validation with the forward event selection from ATLAS~\cite{ATLAS:2020bhl} and is listed in this section.  The second is the event selection which will be described in the following section.  Cuts applied at the generator level are listed in App.~\ref{app:generator}.

We denote $b_1$ and $b_2$ as the leading and subleading $b$-tagged jets which form the highest $p_T$ $b$-jet pair in the events. $j_1$ and $j_2$ refer to the tagging jets, where $j_1$ is the jet with highest $p_T$ and $j_2$ is the sub-leading jet\footnote{Despite the fact that we are looking for a high $p_T$ Higgs, in the VBF channel, the peak sensitivity comes from the range $p_T^h \approx 300~{\rm GeV}$.  Therefore, the analysis can safely resolve the $b$-jets rather than using a boosted analysis looking for a Higgs jet with substructure~\cite{Butterworth:2008iy}.  This was also the case in Ref.~\cite{Araz:2020zyh} where they searched for resolved $\gamma\gamma$ and resolved hadronic $\tau^+\tau^-$.}
\begin{itemize}

    \item $p_T^{b_1} > 85$ GeV and $|\eta^{b_1}| < 2.5$.  The operating point used for $b$-tagging this $b$-jet has a 77\% tagging efficiency.
    
    \item $p_T^{b_2} > 65$ GeV and $|\eta^{b_2}| < 2.5$.  The operating point used for $b$-tagging this $b$-jet has a 85\% tagging efficiency.
    
    \item $p_T^{j_1} > 60$ GeV and $3.2 < |\eta^{j_1}| < 4.5$.\footnote{Note that $j_1$ is always the hardest jet.  An event where, for example, $p_T^{j_1}=90~{\rm GeV}$, $\eta^{j_1}=3.0$, $p_T^{j_2}=80~{\rm GeV}$, $\eta^{j_2}=-4.0$ does not pass the selection.}
    
    \item $p_T^{j_2} > 30$ GeV and $|\eta^{j_2}| < 4.5$.
    
    \item $p_T^{b\bar{b}} > 150$ GeV, where $p_T^{b\bar{b}}$ is the transverse momentum of the $b\bar{b}$ system.
    
\end{itemize}
To validate our analysis, we compare our {\tt VBF} and {\tt QCD} event yields with the ATLAS study~\cite{ATLAS:2020bhl} with the above pre-selection applied. The numerical results are shown in Table~\ref{tab:validation}.  Note that Ref.~\cite{ATLAS:2020bhl} is searching for the Standard Model Higgs signal so their signal corresponds to our {\tt VBF} while their (data-driven) background corresponds to our {\tt QCD}.

\begin{table} [htb!]
\centering
\begin{tabular}{c|c c|c c}
 & {\tt VBF} & ATLAS signal & {\tt QCD} & ATLAS data sidebands \\\hline
after pre-selection & 755.8 & 930.9 & 1,683,155 & 2,584,704 \\ 
\end{tabular}
\caption{Event yields from our simulation for {\tt VBF} and  {\tt QCD} and expected event yields from Ref.~\cite{ATLAS:2020bhl}, at $\sqrt{s} = 13$ TeV with an integrated luminosity of $126~\text{fb}^{-1}$.}
\label{tab:validation}
\end{table}

It is expected that our expected number of events is lower than those from ATLAS because we only generate events at leading order.  Higher-order corrections can account for this difference.  In VBF the $k$-factor at next-to-leading order is known to be $\approx 1.1$ \cite{Greljo:2017spw,Buckley:2021gfw}.  The ratio between the ATLAS event yield and ours is about $1.2$.  We apply an effective $k$-factor of $1.2$ to our {\tt VBF} and {\tt BSM} samples.  In our {\tt QCD} sample we use an effective $k$-factor of $1.5$ which makes our results consistent with the ATLAS event yield.

\subsection*{Event Selection}

In addition to the above pre-selection, we require the events to satisfy following selections which are based on the characteristics of our {\tt BSM} signal, inspired by \cite{Araz:2020zyh}, but tailored to our final state.

Since the VBF signal features a wide rapidity gap between the two tagging jets, events with large separation in pseudorapidity are selected for
\begin{equation}
    \Delta\eta^{j_1,j_2} > 3.
\end{equation}
The two tagging jets tend to be in the forward and backward region, so we require them to be in the opposite directions relative to the beam axis
\begin{equation}
    \eta^{j_1} \cdot \eta^{j_2} < 0.
\end{equation}
Due to this back-to-back nature, the two tagging jets also tend to have large invariant mass.  We select events with
\begin{equation}
    m^{j_1,j_2} > 600~{\rm GeV}.
\end{equation}
To limit additional emissions, we demand the minimum angular separation between any $b$-jet and any tagging jet to be
\begin{equation}
    \Delta R_{\rm min}^{b_i,j_k}> 1.5, \quad\quad i,k = 1, 2.
\end{equation}
Furthermore, we make use of the azimuthal angular separation between the $b$-jet pair and the tagging jet pair, and we require
\begin{equation}
    \Delta \Phi^{b_1 b_2,j_1 j_2} > 1.5,
\end{equation}
where $\Delta \Phi^{b_1 b_2,j_1 j_2}$ is the difference in azimuthal angle between the $b_1 b_2$ system and the $j_1 j_2$ system.

Since the QCD background tends to have jets with a relatively large azimuthal angular separation, we select events such that
\begin{equation}
    \Delta \Phi^{j_1,j_2} < 2.
\end{equation}
In addition, because of the presence of the Higgs boson in the central region, we require it to lie between the two tagging jets
\begin{equation}
    y_{\rm min}^{j_1, j_2} < y^h < y_{\rm max}^{j_1,j_2},
\end{equation}
where $y$ is the the rapidity of the jet or the reconstructed Higgs boson.

Finally, we cut on the invariant mass of the $b\bar{b}$ system.  Due to the mass resolution for $b\bar{b}$, we require a relatively wide mass window around the Higgs boson mass
\begin{equation}
    115~{\rm GeV} < m^{b_1 b_2} < 135~{\rm GeV}.
\end{equation}
These cuts effectively distinguish {\tt VBF} from {\tt QCD}.  Our goal, however, is to identify the contribution of new physics to {\tt BSM} from both {\tt QCD} and from {\tt VBF}.  

\begin{figure} [tb!]
  \begin{center}
  \includegraphics[width=0.45\textwidth]{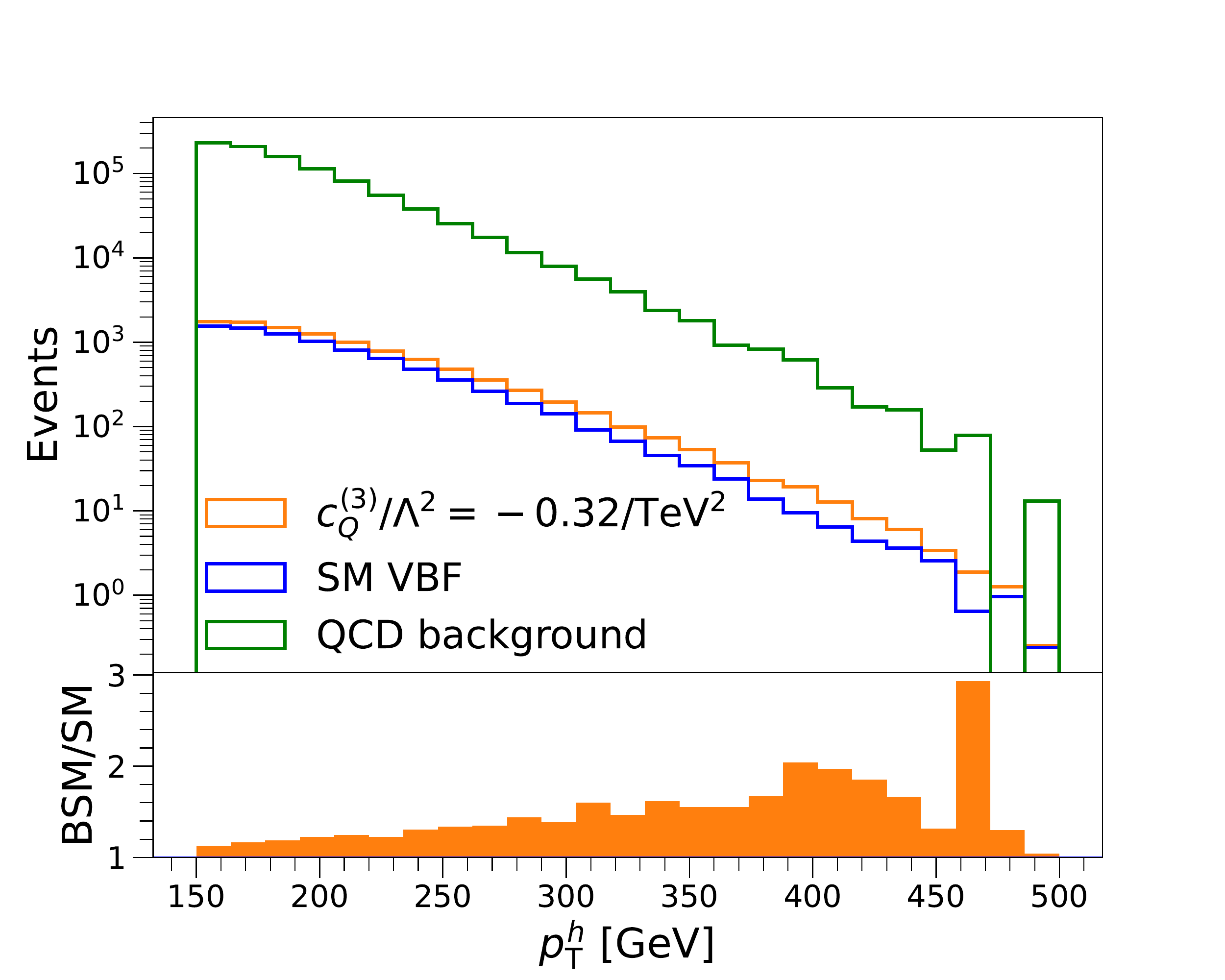} \quad\quad
  \includegraphics[width=0.45\textwidth]{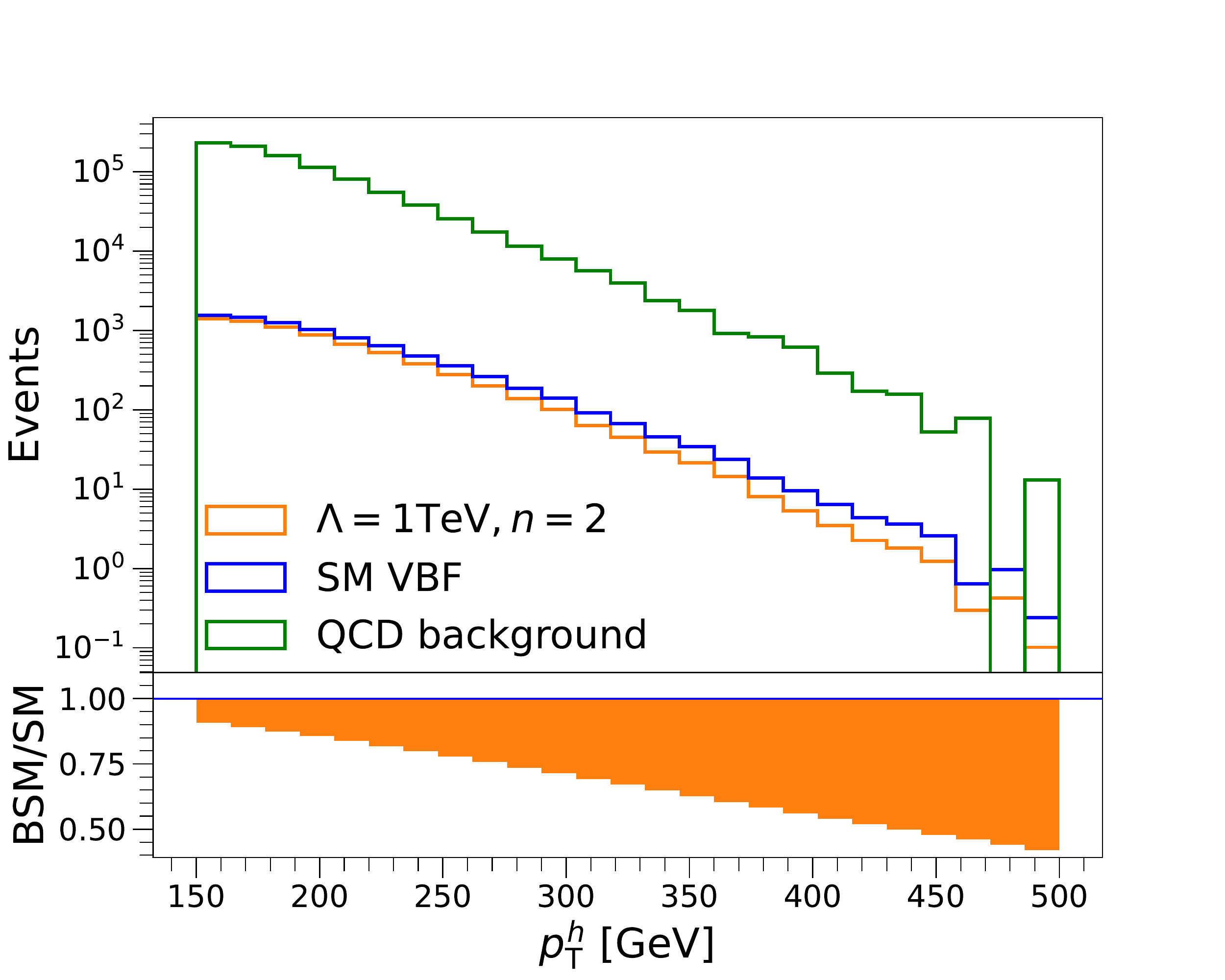}
  \caption{Distribution of the Higgs $p_T$ in {\tt QCD}, {\tt VBF}, and {\tt BSM}.  On the left, $c_Q^{(3)} / \Lambda^2$ is set to $-0.32~{\rm TeV}^{-2}$, while on the right, a form factor with $n=2$ and a scale $\Lambda=1$ TeV is used.  The integrated luminosity is $\mathcal{L} = 3~{\rm ab}^{-1}$. }
  \label{fig:distributionHiggsPT}
  \end{center}
\end{figure}

To distinguish {\tt BSM} from {\tt VBF}, consider Fig.~\ref{fig:distributionHiggsPT} which shows the distribution of the Higgs $p_T$ (after both pre-selection and event selection) with a non-zero value for $c_Q^{(3)}$ (left) and including the $n=2$ form factor (right).  We see that this is a powerful discriminant to separate our signal from the backgrounds.  Note that new physics from dimension-6 operators can cause an increase or decrease in the number of expected events because the leading effect is the interference between new physics and the SM.

\section{Results}
\label{sec:results}

To derive the sensitivity to a new physics scale at the LHC, we adopt the usual statistical significance, 
\begin{equation}
    \frac{S}{\sqrt{B}} = \frac{|N_{\tt BSM} - N_{\tt VBF}|}{\sqrt{N_{\tt VBF} + N_{\tt QCD}}},
\end{equation}
where $N_i = \mathcal{L} \cdot \sigma_i$ denotes the expected numbers of events after event selections with luminosity $\mathcal{L}$ for sample $i=\{ {\tt BSM}, {\tt VBF}, {\tt QCD} \}$.  

The contribution of new physics can result in either an enhancement or a suppression relative to the SM VBF yield. In particular, the form factor causes a suppression in the signal where the effect becomes more prominent when $\Lambda$ is small.  The effect of the dimension-6 operators depends on the size of the Wilson coefficient and the sign of the interference with the SM.  In the case that $N_{\tt BSM}$ is less than $N_{\tt VBF}$, we take the absolute value to ensure a positive significance.

\begin{figure} [H]
  \begin{center}
  \includegraphics[width=0.49\textwidth]{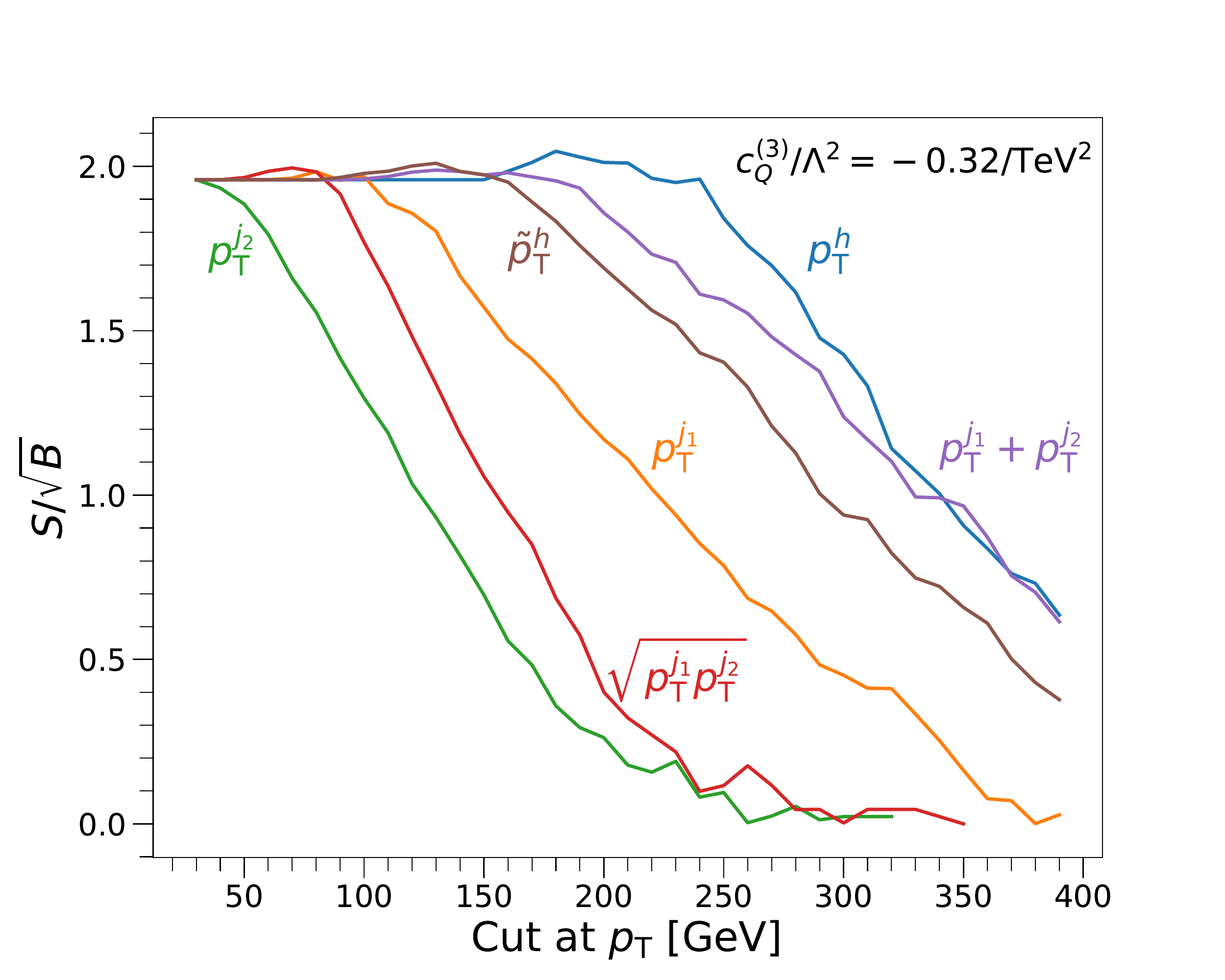} 
  \includegraphics[width=0.49\textwidth]{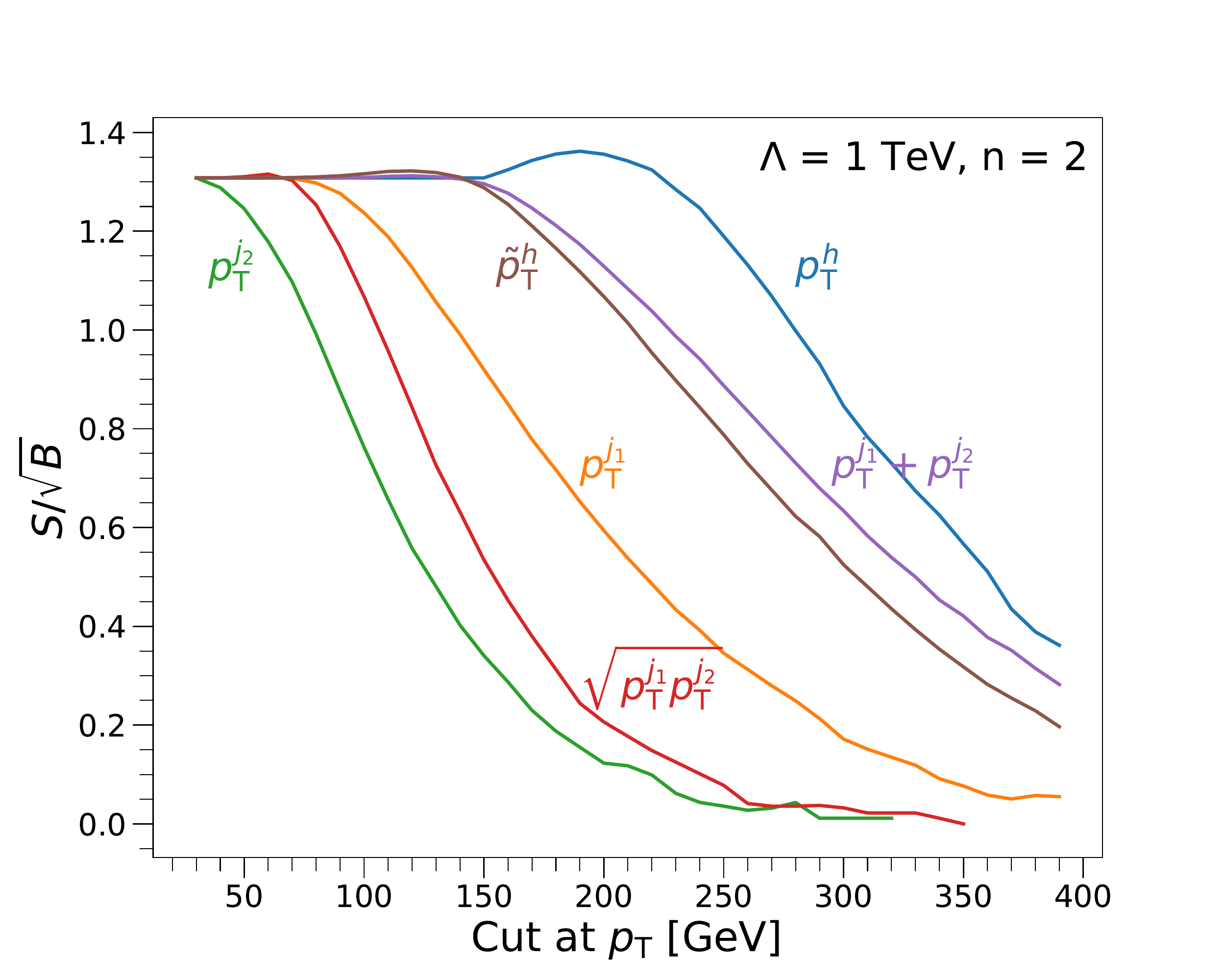}
  \caption{Search significance as a function of cut on Higgs $p_T$-like variables for coefficient $c_Q^{(3)}$ (left) and for the $n=2$ form factor (right).  Variables used are described in Sec.~\ref{sec:kinematics}.  The integrated luminosity is $\mathcal{L} = 3~{\rm ab}^{-1}$. 
  \label{fig:significanceDim6} }
  \end{center}
\end{figure}

In Fig.~\ref{fig:significanceDim6}, we show how selecting different observables at different thresholds affects the significance achieved at high luminosity LHC assuming $\mathcal{L} = 3~\text{ab}^{-1}$, for the case where $c_Q^{(3)}/\Lambda^2 = -0.32~{\rm TeV}^{-2}$ (left) and the $n=2$ form factor with $\Lambda=1~{\rm TeV}$ (right).

From Fig.~\ref{fig:significanceDim6} we find the transverse momentum of the Higgs boson to be the most powerful observable to distinguish new physics from SM, as we discussed in Sec.~\ref{sec:simulation}.   The other variables we consider are $p_T^{j_1} + p_T^{j_2}$, $\tilde{p}_T^h$ (see Eq.~\eqref{eq:kinematics}), $p_T^{j_1}$, $\sqrt{p_T^{j_1} p_T^{j_2}}$, and $p_T^{j_2}$.  The tagging jets, in particular the leading jet, can be useful to probe the high momentum-exchange region of VBF.  The subleading jet alone, however, does not correlate strongly with the Higgs $p_T$ and is less useful.

\begin{figure} [H]
  \begin{center}
  \includegraphics[width=0.80\textwidth]{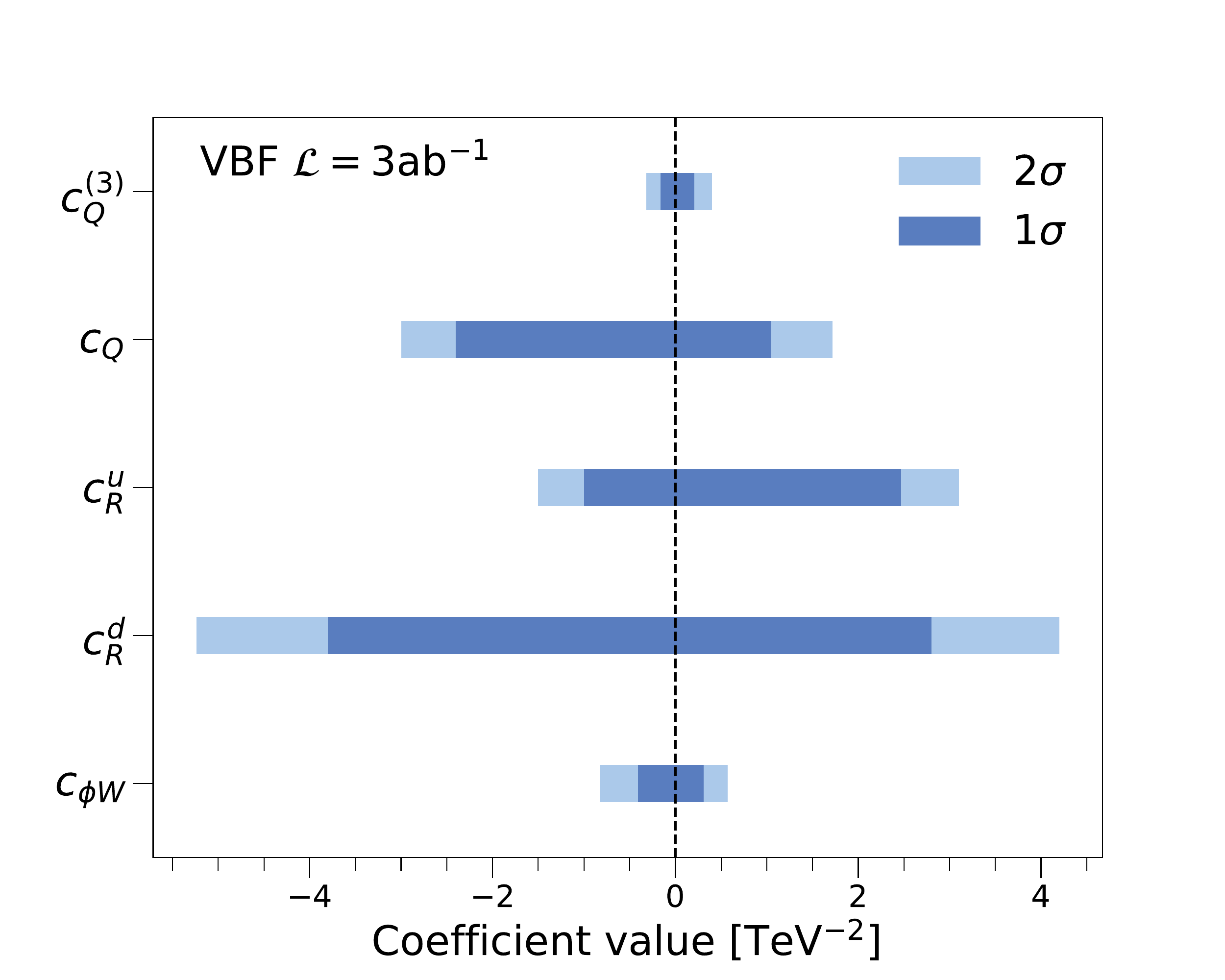}
  \caption{Limits on $c_i / \Lambda^2$ values of dimension-6 operators with an integrated luminosity of $3~{\rm ab}^{-1}$. }
  \label{fig:limitsDim6}
  \end{center}
\end{figure}

Consequently, to compute the new physics reach, we use $p_T^h$ as the final discriminant and select the threshold that optimizes the significance.  The optimal value does not differ much between operators and is around 250 GeV.  We do not apply an upper cut on $p_T^h$ (as would be safer for EFT validity), however, this has very little effect on the results since the sensitivity is driven by the bins near $p_T^h \approx 300~{\rm GeV}$~\cite{Araz:2020zyh}. We have explicitly verified that bins above 400 GeV only affect our sensitivity at the percent level.   The corresponding $1\sigma$ and $2\sigma$ constraints on the dimension-6 operators are shown in Fig.~\ref{fig:limitsDim6}.  The asymmetric limit between positive and negative coefficients is expected because the leading contribution to the rate is the interference term between the SM and new physics.

The 95\% C.L. limits are given numerically in Table~\ref{tab:resultsDim6}.  In the top row, we give the limits on $c_i / \Lambda^2$ for each operator, one at a time.  The VBF search is most sensitive to the $\cO_Q^{(3)}$ operator, as expected~\cite{Araz:2020zyh}.  Fixing $c_i = \pm 1$ allows us to translate the bound into a limit on the scale $\Lambda$ which we show in the middle and bottom rows of the table, respectively.  Compared to the combination of the $\gamma\gamma$ and $\tau^+\tau^-$ decay channels, we find the $b\bar{b}$ channel to be roughly a factor of $2$ less sensitive~\cite{Araz:2020zyh} (in comparing the coefficients).

\begin{table} [H]
\begin{center}
\begin{tabular}{c|c c c c c}
\hline \hline
 & $c_Q^{(3)}$ & $c_Q^{(1)}$ & $c_u$ & $c_d$ & $c_{\phi W}$  \\\hline
$c_i/\Lambda^2~[\text{TeV}^{-2}]$ & [$-0.32$, 0.40] & [$-3$, 1.72] & [$-1.5$, 3.1] & [$-5.24$, 4.20] & [$-0.82$, 0.57]  \\ \hline
$\Lambda$ [TeV] ($c_i = +1$)  & 1.6 & 0.76 & 0.57 & 0.49 & 1.3 \\ \hline 
$\Lambda$ [TeV] ($c_i = -1$)  & 1.8 & 0.58 & 0.82 & 0.44 & 1.1\\ \hline \hline
\end{tabular}
\caption{95\% C.L.~limits for dimension-6 operators with an integrated luminosity of $3~{\rm ab}^{-1}$.}
\label{tab:resultsDim6}
\end{center}
\end{table}

We can also compare our reach for $\cO_Q^{(3)}$ with the reach in the diboson channel.  Ref.~\cite{Franceschini:2017xkh} studied the channel $q\bar q \to W^* \to W(\ell\nu)Z(\ell\ell)$ and found the reach to be roughly $30$ times better.\footnote{Note that Ref.~\cite{Franceschini:2017xkh} sets limits on dimensionful coefficients $a_i$, which correspond to our parametrization via $a_i = 4 c_i / \Lambda^2$.}  While diboson is an experimentally cleaner channel than VBF, it is interesting to note that the rate for longitudinally-polarized vector bosons is heavily suppressed relative to the rate for transversely-polarized vector bosons~\cite{Contino:2010mh}.  One consequence of this is that the limits on $c_{\phi W}$ in VBF are only a factor of $2$ weaker than on $c_Q^{(3)}$, while in diboson the limits are a factor of $24$ weaker~\cite{Bishara:2020vix}.

\begin{figure} [tb!]
  \begin{center}
  \includegraphics[width=0.75\textwidth]{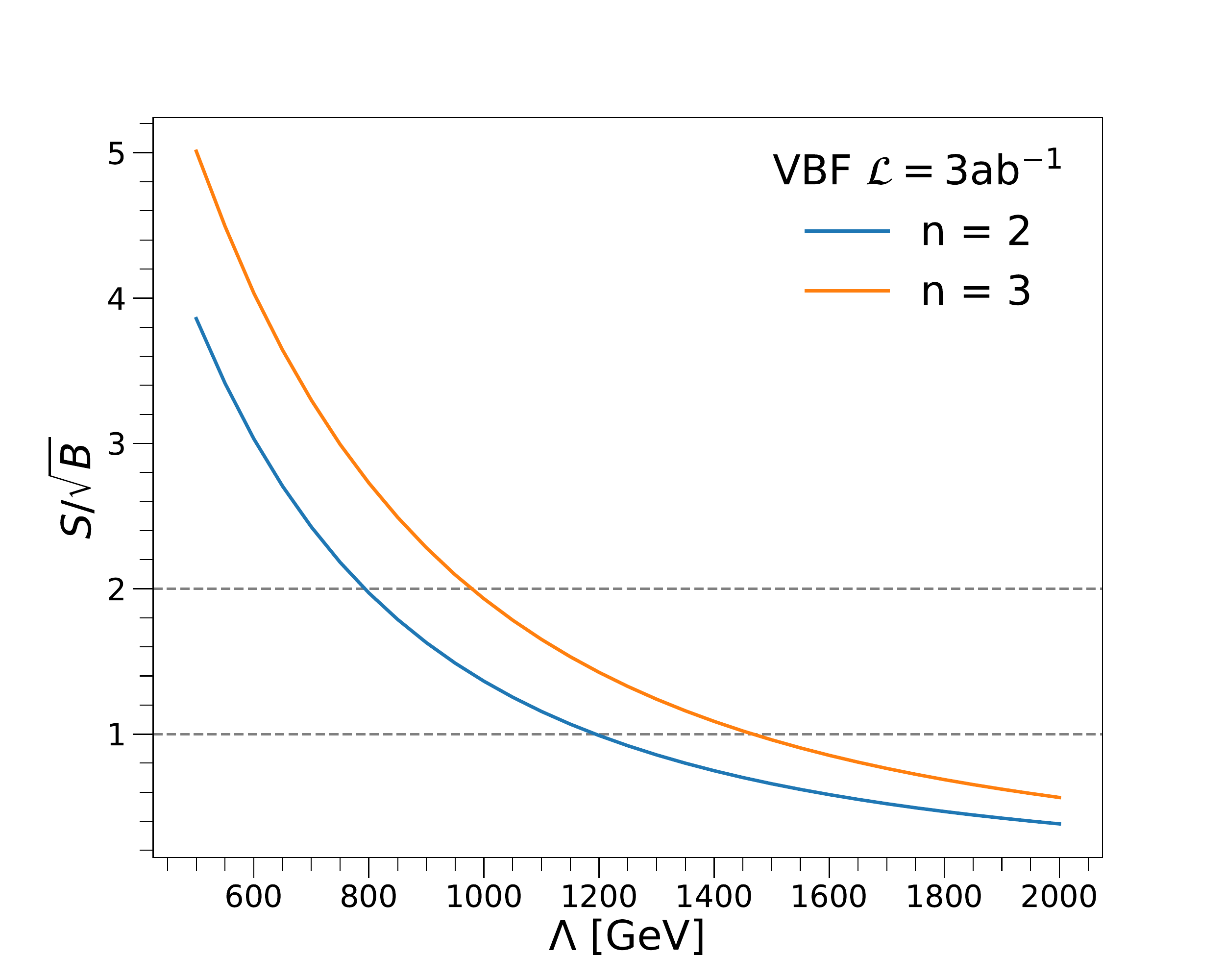}
  \caption{Search significance as a function of the scale $\Lambda$ for the form factor Eq.~\eqref{eq:formFactor} for $n=2$ (blue) and for $n=3$ (orange).  An integrated luminosity of $3~{\rm ab}^{-1}$ is used.}
  \label{fig:formFactorLimits}
  \end{center}
\end{figure}

Figure~\ref{fig:formFactorLimits} shows the search significance as a function of $\Lambda$ for the form factor, Eq.~\eqref{eq:formFactor}, with $n=2$ and $n=3$.  Since $n=3$ has a steeper suppression than $n=2$, the significance for fixed $\Lambda$ is larger.  The 95\% C.L. limits are 0.8 TeV ($n=2$) and 1 TeV ($n=3$).  There are no other searches for this form factor of $VVh$ interactions, but for the analogous form factor applied to the $t\bar{t}h$ interaction, the projected limits for $n=2$ are $\Lambda=0.8~{\rm TeV}, 1.5~{\rm TeV}, 2.1~{\rm TeV}$, respectively for the channels $h^* \to ZZ \to \ell\ell\ell\ell$, $h^* \to ZZ \to \ell\ell\nu\nu$, and $t\bar{t}h$~\cite{Goncalves:2020vyn,MammenAbraham:2021ssc}.  For $n=3$ they are $\Lambda=1.1~{\rm TeV}, 2.1~{\rm TeV}, 2.7~{\rm TeV}$ in the same channels.

\begin{figure}[H]
  \begin{center}
  \includegraphics[width=0.75\textwidth]{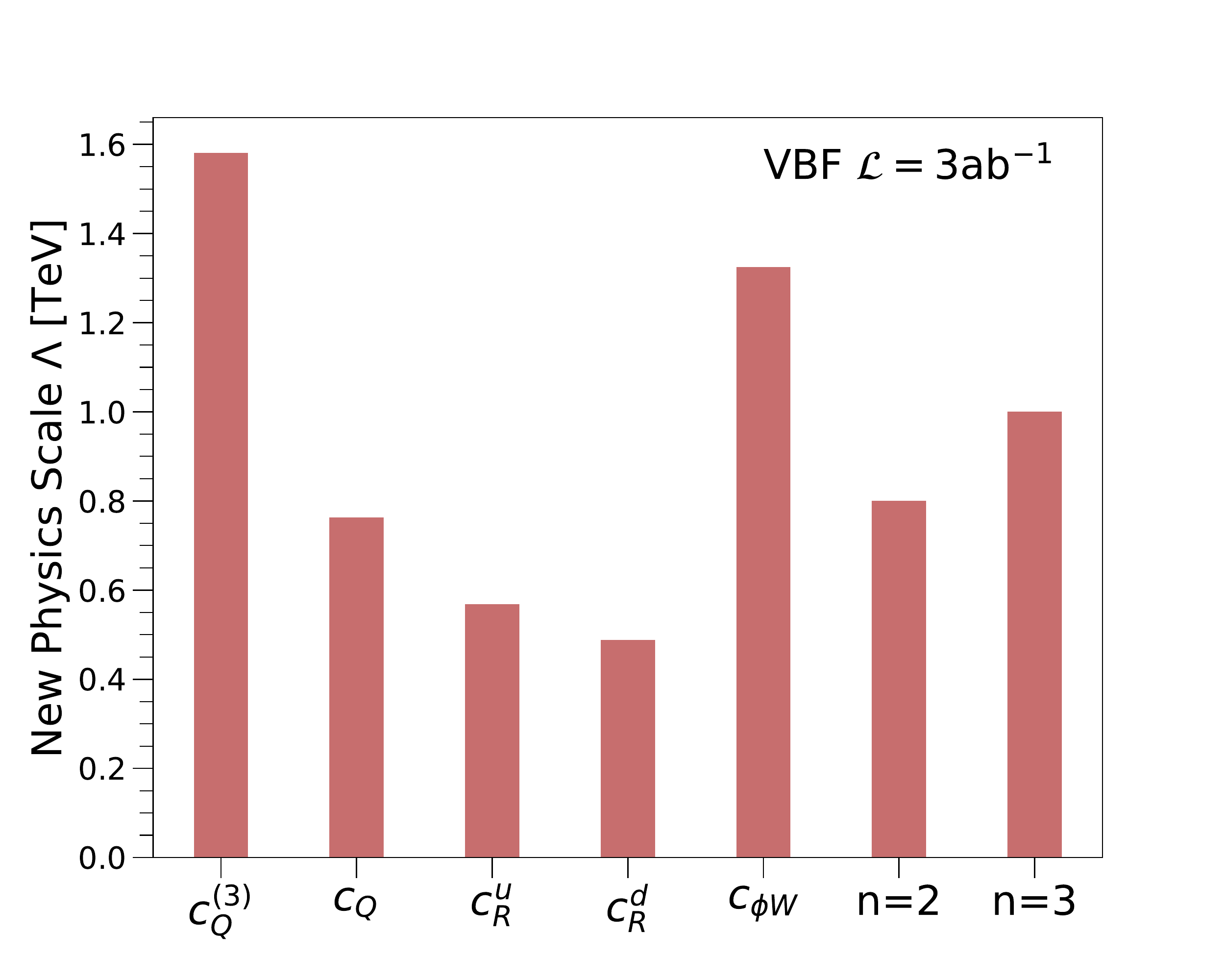}
  \caption{Limits on the scale of new physics $\Lambda$ at 95\% C.L. for dimension-6 operators (assuming positive coefficients) and the form factor with $n=2$ and $n=3$ with an integrated luminosity of $3~{\rm ab}^{-1}$. }
  \label{fig:limitsAll}
  \end{center}
\end{figure}

Finally, we summarize all new physics results together in Fig.~\ref{fig:limitsAll}.  The strongest limits are set on the $c_Q^{(3)}$ coefficient, as the VBF rate depends mostly strongly on its value~\cite{Araz:2020zyh}.  Reasonable sensitivity is still maintained if only one of the other coefficients $c_Q$, $c_R^u$, $c_R^d$ is present.  This is because in VBF there are always contributions from both $Z$ and $W$ bosons. 

\section{Conclusions}
\label{sec:conclusions}

In this paper we study the phase space region of VBF where the Higgs $p_T$ is large.  In this region, there is sensitivity to new physics coming from high scales, complementary to off-shell Higgs boson physics.  We study the impact of both dimension-6 operators and a nuclear-physics-inspired form factor.

We explore the $h(b\bar{b})jj$ channel where the Higgs decays on-shell to $b\bar{b}$ at a high transverse momentum, in the hope to probe the physics at high scales. The main backgrounds are the SM contribution to VBF and QCD where the $b\bar{b}$ is from the splitting of a gluon.  For dimension-6 operators, the reach for new physics is up to 1.8 TeV at 95\% C.L.  This reach is a little weaker than previous searches in the $\gamma\gamma$ and $\tau^+\tau^-$~\cite{Araz:2020zyh} where the decrease is due to the reduced Higgs mass resolution and the much larger backgrounds in the $b\bar{b}$ channel.

Ultimately to maximize our reach for new physics the next step would be to combine all existing channels.  Additionally, it would be useful to include additional decay channels of the Higgs such as $W(\ell\nu)W(\ell\nu)$.  As we have discussed, for such a channel with invisible particles, one can cut on $p_T^{j_1}+p_T^{j_2}$, since $p_T^h$ would be unavailable, and maintain sensitivity to the high momentum exchange region of phase space.  With enough data, it would also be interesting to do a simultaneous fit to VBF and diboson data to fully characterize the full space of high energy primary operators.  Combining as many channels as possible is especially critical in VBF due to the suppression in the distribution of longitudinal vectors~\cite{Contino:2010mh}.  The contribution of the $b\bar{b}$ channel is an indispensable component.

\begin{acknowledgments}
The authors would like to thank Joseph Boudreau for computing assistance and Gauthier Durieux for consultation on the FeynRules model {\tt SMEFTatNLO}.
This work was supported by the U.S.~Department of Energy under grant No.~DE-FG02-95ER40896 and by the PITT PACC.
\end{acknowledgments}

\appendix
\section{Generator-Level Cuts}
\label{app:generator}

In each sample the final state at parton-level includes two $b$-quarks and two light flavor quarks.  The $b$-quarks are labelled as $b_1$ and $b_2$, in descending order of $p_T$, while the light flavor quarks are labelled as $j_1$ and $j_2$ following the same scheme.
\begin{itemize}

    \item $p_T^{j_1} > 40$ GeV, $p_T^{j_2} > 10$ GeV.
    
    \item $p_T^{b_1} > 65$ GeV, $p_T^{b_2} > 45$ GeV.
    
    \item $3 < |\eta^{j_1}| < 5$, $|\eta^{j_2}| < 5$, $|\eta^{b_1}| < 3$, $|\eta^{b_2}| < 3$
        
    \item $|\eta^{j_1} - \eta^{j_2}| > 2.5$.
    
    \item $\Delta R^{j_1,j_2}>0.4$, $\Delta R^{b_1,b_2}>0.4$, $\Delta R^{b,j} > 0.4$, where $\Delta R^{b,j}$ refers to any pair of $b$-quark and light flavor quark and $\Delta R^{a,b} = \sqrt{(\eta^a - \eta^b)^2 + (\phi^a-\phi^b)^2}$.
    
    \item $p_T^{b\bar{b}} > 120$ GeV, where $p_T^{b\bar{b}}$ is the transverse momentum of the $b\bar{b}$ system.
    
    \item $m^{jj} > 550$ GeV where $m^{jj}$ is the invariant mass of the dijet system.
    
    \item 110 GeV $< m^{b\bar{b}} < 140$ GeV, where $m^{b\bar{b}}$ is the invariant mass of the $b\bar{b}$ system.
    
\end{itemize}

\begin{figure} [htb!]
  \begin{center}
  \includegraphics[width=0.75\textwidth]{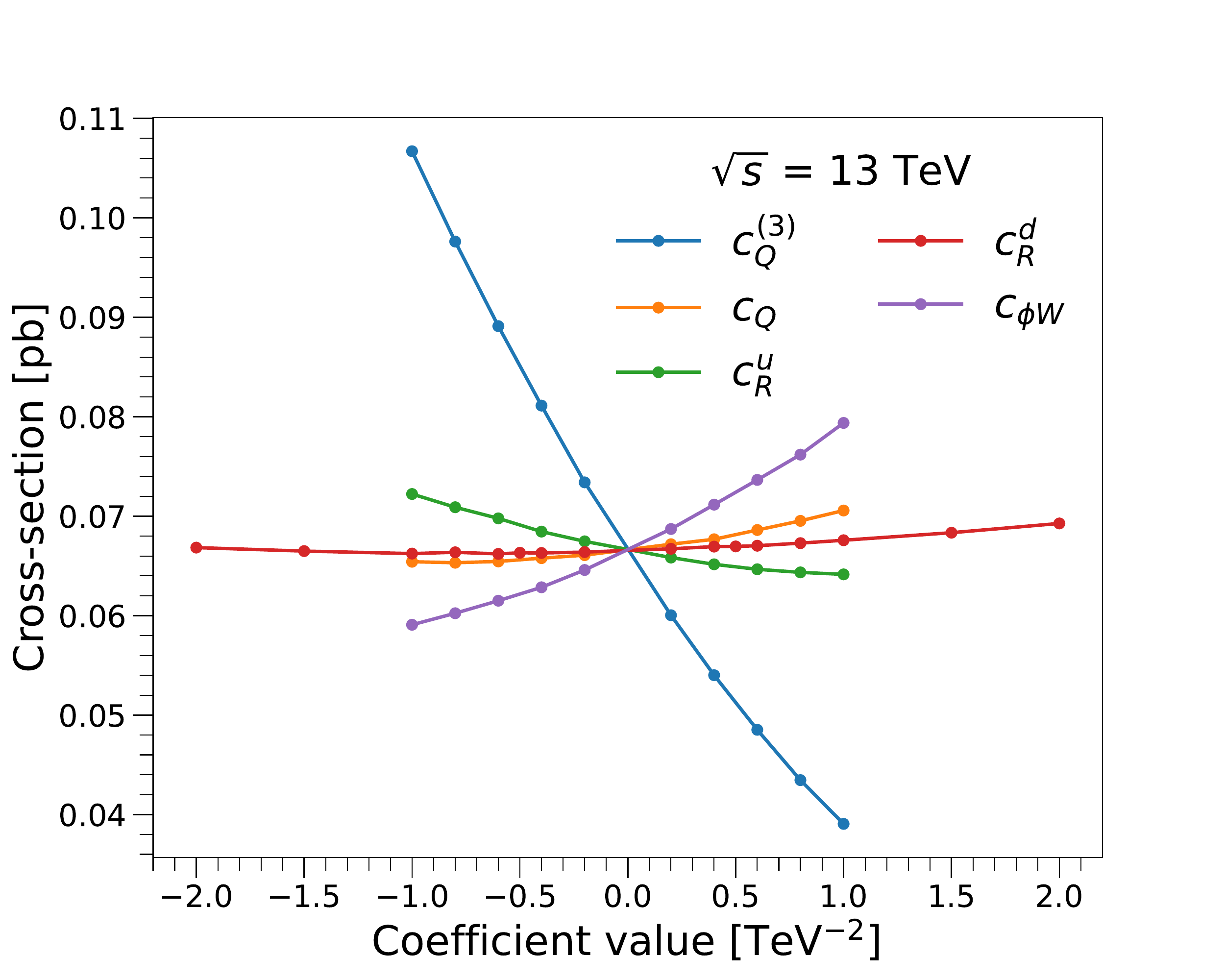}
  \caption{Cross section for {\tt BSM} sample where the coefficient of a single operator is varied. }
  \label{fig:xsecVsCoefficient}
  \end{center}
\end{figure}

Already with these generator-level cuts, the impact of dimension-6 is observable.  Figure~\ref{fig:xsecVsCoefficient} shows the cross section of {\tt BSM} where the coefficient of a single dimension-6 operator coefficient is varied.

\bibliographystyle{JHEP}
\bibliography{refs}
\end{document}